\documentclass[12pt,notitlepage]{article}
\usepackage{amssymb}

\usepackage{graphicx}
\usepackage{amsmath}

\input{tcilatex}

\begin{document}

\begin{center}
{\LARGE IS MASS QUANTIZED?}

\bigskip

\bigskip

\bigskip

Paul S. Wesson$^{1}$

\bigskip

\bigskip

\bigskip
\end{center}

\begin{enumerate}
\item Dept. of Physics, University of Waterloo, Waterloo, Ontario \ \ N2L
3G1 \ Canada
\end{enumerate}

Keywords: \ Cosmological Constant, Quantization, Higher-Dimensional Field
Theory

PACs: 03.70.+k, 11.90.+t, 11.10Kk

Correspondence: post to address (1) above, phone (519)885-1211 x2215, fax
(519)746-8115, email wesson@astro.uwaterloo.ca

\pagebreak

\bigskip \underline{{\Large Abstract}}

The cosmological constant combined with Planck's constant and the speed of
light implies a quantum of mass of approximately $2\times 10^{-65}$ g. \
This follows either from a generic dimensional analysis, or from a specific
analysis where the cosmological constant appears in 4D spacetime as the
result of a dimensional reduction from higher dimensional relativity (such
as 5D induced-matter and membrane theory). \ In the latter type of theory,
all the particles in the universe can be in higher-dimensional contact.

\section{\protect\underline{Introduction}}

\ \ Observations of the cosmic microwave background, the dynamics of
galaxies and the gravitational lensing of remote astronomical sources such
as quasars imply that 99\% of the material in the universe is dark matter.$%
^{1}$ \ Of this, a significant fraction appears to be due to the equivalent
density of the ``vacuum''. \ This in general relativity is measured by $%
\Lambda c^{2}\diagup 8\pi G$ where $\Lambda $ is the cosmological constant, $%
c$ is the speed of light and $G$ is the gravitational constant.$^{2}$ \ The
size of the first parameter is $\Lambda \simeq 3\times 10^{-56}$ cm$^{-2}$
approximately. $^{3,4}$ \ (Equivalently, the observed universe has an
intrinsic length scale of order 10$^{28}$ cm, or an age of order 10$^{10}$
years, with a vacuum density of order 10$^{-29}$ g cm$^{-3}$.) \ It is now
widely believed that the cosmological constant of general relativity is a
parameter which is derived from the reduction to 4D of higher-dimensional
theories whose motivation is to unify gravity with the interactions of
particle physics.$^{2}$ \ These include 10D supersymmetry, 11D supergravity
and 26D string theory. \ They should provide a natural place for the quantum
of action as measured by Planck's constant $h$. \ The basic extension of 4D
Einstein theory and the low-energy limit of higher-D theories is the modern
incarnation of (non-compact) 5D Kaluza-Klein theory. \ This has been
intensively studied recently, under the names induced-matter theory$^{5,6}$
and membrane theory.$^{7,8}$ \ Both admit a fifth force which may be
relevant to the interactions of particles$^{9,10}$ and both allow us to view
massive particles in spacetime as massless or photon-like in the larger
manifold $^{11,12}$ (i.e., timelike 4D paths may be viewed as 5D null
geodesics). \ Also, the field equations of both versions of 5D relativity
have recently been shown to be equivalent$^{13}$. In regard to comparison of
the 5D field equations with observations, Campbell's theorem guarantees an
embedding of the 4D Einstein equations and their Newtonian limit$^{14}$, and
an explicit calculation shows that the classical tests of relativity are
satisfied$^{15}$. \ In comparison with cosmological data, Campbell's theorem
means that the standard models are recovered, and there are explicit 5D
solutions which represent the present universe$^{2,16}$ as well as the early
universe with inflation$^{2,17}$ and an effective cosmological constant.

These results mean that we are now in a position to take a fresh look at the
constants $c$, $G$, $h$ and $\Lambda $. \ In what follows, we will do this
first in a generic sense, using only dimensional analysis; and then we will
collect some technical results from 5D relativity to give a more specific
account. \ Both approaches will be seen to imply that mass is quantized at
the level of approximately $2\times 10^{-65}$g.

\section{\protect\underline{Cosmological Constant and Mass Scales}}

\qquad In this section, we will use the fundamental parameters and then a
more detailed analysis to show that the cosmological constant implies two
distinct mass scales. \ Of these, one is a minimum \ and so defines a
quantum of mass. \ Some of what follows may be familiar, but in addressing
such a fundamental issue we wish to ensure that the traditional approach and
the new one are compatible.

Dimensional analysis is an elementary group-theoretic technique$^{18-22}$. \
It is related to the fact that the equations of physics are homogeneous in
their physical dimensions, which implies that they can be written in terms
of dimensionless quantities if we so wish (this is the basis of modelling
theory). \ The technique involves the ability to transform quantities of
different physical types to ones of the same physical dimensions using the
constants (this is the basis of using $x^{0}\equiv ct$ to transform the time
to a length coordinate in relativity). \ The technique also implies the
freedom to choose \underline{units}, which in mechanics means fiducial
values for the base physical dimensions $M$, $L$, $T$ of mass, length and
time (this is the basis of the convenient choice $c=1$, $G=1$ in
relativity). \ However, dimensional analysis is a generic technique, without
detailed knowledge of the underlying theory to which it is applied (which is
why it fails to determine dimensionless factors such as integers and $\pi $%
). \ Also, it is problematic in application when the constants have physical
dimensions which ``overlap'' or are degenerate$^{21}$. \ This is the case
presently being encountered in cosmology with the recognition of $\Lambda $
as a fundamental constant. \ To illustrate what is involved here, we simply
have to realize that from the 4 constants $c$, $G$, $h$ and $\Lambda $\ we
can form \underline{two} different masses%
\begin{equation}
m_{P}\equiv \left( h\diagup c\right) \left( \Lambda \diagup 3\right)
^{1/2}\simeq 2\times 10^{-65}\,\text{g\ \ \ \ \ \ \ }
\end{equation}%
\begin{equation}
m_{E}\equiv \left( c^{2}\diagup G\right) \left( 3\diagup \Lambda \right)
^{1/2}\simeq 1\times 10^{56}\,g\;\;\;\;\;.
\end{equation}%
Here the two masses are relevant to quantum and gravitational situations,
and so may be designated by the names Planck and Einstein respectively. \
[To avoid confusion, it can be mentioned that the mass $m_{PE}\equiv \left(
hc\diagup G\right) ^{1/2}\simeq 5\times 10^{-5}$ g which is sometimes called
the Planck mass does \underline{not} involve $\Lambda $\ and \underline{mixes%
} $h$ and $G$. \ From the viewpoint of higher-dimensional field theory as
outlined below, this is equivalent to mixing gauges and is ill-defined,
possibly explaining why this mass is not manifested in nature$^{20}$.] \ The
mass (2) is straightforward to interpret: it is the mass of the observable
part of the universe, equivalent to 10$^{80}$ baryons of 10$^{-24}$ g each.
\ The mass (1) is more difficult to interpret: it is the mass of a quantum
perturbation in a spacetime with very small local curvature, measured by the
astrophysical value of $\Lambda $ as opposed to the one sometimes inferred
from the zero-point or vacuum fields of particle interactions. \ We are
fully cognizant of this mismatch, which is commonly called the
cosmological-constant problem $^{1,22-26}$. \ [Its essence is that if one
believes $m_{PE}$ to be a physical mass scale then $\Lambda $\ in (1) has to
be larger than that in (2) by 10$^{120}$ or so.] \ However, in our approach
the cosmological-constant problem becomes moot, because even if $\Lambda $
were larger in localized regions of space$^{2}$ or in the early universe$^{4}
$, its astrophysical value is still a minimum and so the mass (1) is still
the smallest one possible.

Higher-dimensional field theory provides not only a rationale for the $%
\Lambda $ we measure in spacetime but also an account of 4D dynamics (based
on solutions of the field equations and the equations of motion). \ In the
basic 5D theory, the ``separation'' (squared) between two nearby points is
given by the line element $dS^{2}=g_{AB}dx^{A}dx^{B}$, which contains that
of general relativity $ds^{2}=g_{\alpha \beta }dx^{\alpha }dx^{\beta }$. \
(Here $A,B=0,123,4$ and $\alpha ,\beta =0,123$ for coordinates $x^{0}=ct$, $%
x^{123}=xyz$ and $x^{4}=l$. \ There is a summation over indices repeated in
the metric tensor and the coordinate elements.) \ For the induced-matter
version of the theory most work has been done using the canonical form of
the metric$^{6}$, while for the membrane version most has been done using
the warp form of the metric$^{7}$. \ As noted above, the two versions of 5D
relativity are mathematically equivalent at a general level. \ However, the
results may not be physically equivalent, because both approaches depend on
a choice of coordinates or gauge. \ [This is because the 5D group of
transformations $x^{A}\rightarrow \overline{x}^{A}\left( x^{B}\right) $ is
wider than the 4D group $x^{\alpha }\rightarrow \overline{x}^{\alpha }\left(
x^{\beta }\right) $, so spacetime physics can change under an $l$-dependent
change of gauge: see refs. 2, 11, 20.] With respect to this and a wish to
understand the masses (1), (2) noted above, it is instructive to introduce a
new gauge. \ We call this the Planck gauge. \ And for reasons which will
become apparent, we rename the canonical frame the Einstein gauge. \ The two
are specified respectively by

\begin{equation}
dS^{2}=\left( L\diagup l_{P}\right) ^{2}ds^{2}-\left( L\diagup l_{P}\right)
^{4}dl_{P}^{2}
\end{equation}%
\begin{equation}
dS^{2}=\left( l_{E}\diagup L\right)
^{2}ds^{2}-dl_{E}^{2}\;\;\;\;\;\;.\;\;\;\;\;\;
\end{equation}%
Here $L$ is a constant length introduced to give consistency of physical
dimensions. \ However, it turns out to have great relevance to our present
discussion, because a reduction of the field equations in 5D to their
counterparts in 4D (which include Einstein's equations) shows that $L$ is
related to the cosmological constant via $\Lambda =3\diagup L^{2}$ (see
refs. 2, 6, 17, 23). Thus cosmological data imply $L\simeq 1\times 10^{28}$
cm. \ The extra coordinates $l_{P}$, $l_{E}$ in (3), (4) may be shown by
another reduction of the equations of motion to be related to the Compton
wavelength and Schwarzschild radius of a test particle of mass $m$ via $%
l_{P}=h\diagup mc$ and $l_{E}=Gm\diagup c^{2}$ (see refs. 2, 20, 27; this
can be appreciated directly by noting that with these identifications, the
first parts of the elements of the 5D actions specified by $dS$ involve the
elements of the 4D action $mcds$ specified by the proper time $ds$). \ The
two gauges just stated are related by the simple coordinate transformation $%
l_{E}=L^{2}\diagup l_{P}$, which can be used to go between them. \ They are
not arbitrary, of course, but chosen with care. \ They lead to dramatic
simplifications in the underlying field equations and equations of motion,
and represent the most convenient way to embed the physics of 4D spacetime
in a 5D manifold.

To illustrate the physics inherent in (3) and (4), let us recall that
particles travelling on paths in 4D with $ds^{2}\geq 0$ can be regarded as
travelling on null geodesics in 5D with $dS^{2}=0^{11,12}$. \ The last
statement means that, in some sense, particles are in causal contact in 5D.
\ (They are analogous to photons in 4D, which can be viewed as connecting
events which are separated in ordinary 3D space.) \ This condition with (3)
means that the latter can be rewritten and integrated to yield 
\begin{equation}
\int d\left( L\diagup l_{P}\right) =\left( 1\diagup h\right) \int mcds\,\ \
\ .
\end{equation}%
Here we know that the conventional action is quantized and equal to $nh$
where $n$\ is an integer. \ Thus $L\diagup l_{P}=n$. \ This says that the
Compton wavelength of the particle cannot take on any value, but is
restricted by the typical dimension of the (in general curved) spacetime in
which it exists. \ Putting back the relevant parameters, the last relation
says that $m=\left( nh\diagup c\right) \left( \Lambda \diagup 3\right)
^{1\diagup 2}$. \ For the groundstate with $n=1$, there is a minimum mass $%
m_{P}=\left( h\diagup c\right) \left( \Lambda \diagup 3\right) ^{1\diagup
2}\simeq 2\times 10^{-65}$ g. \ This is the same as (1) above.

A similar procedure to that of the preceding paragraph can be followed for
the Einstein gauge (4). \ However, a notable difference occurs in that the
relation analogous to (5) is now%
\begin{equation}
\int \left( L\diagup l_{E}\right) dl_{E}=\int ds\;\;\;\;.
\end{equation}%
Here we do not have any evidence that the line element by itself is
quantized, so the discreteness which is natural for the Planck gauge does
not carry over to the Einstein gauge. \ However, in the Planck gauge the
condition $L\diagup l=n$ could have been used to reverse the argument and
deduce the quantization of the action from the quantization of the fifth
dimension, implying that the latter may be the fundamental assumption. \ Let
us take this in the form $L\diagup l_{E}=n$. \ [This by (6) then implies $%
dl\diagup ds=1\diagup n$, which also by (5) holds in the Planck gauge. \ The
velocity in the fifth dimension is related to electric charge in certain
approaches to 5D relativity, including the early one of Klein$^{2}$.] \ Then
putting back the relevant parameters, we obtain $m_{E}=\left( c^{2}\diagup
nG\right) \left( 3\diagup \Lambda \right) ^{1\diagup 2}$. \ For the
groundstate with $n=1$, there is a maximum mass $m_{E}=\left( c^{2}\diagup
G\right) \left( 3\diagup \Lambda \right) ^{1\diagup 2}\simeq 1\times 10^{56}$
g. \ This is the same as (2) above.

In summary, astrophysical data indicate that we should add the cosmological
constant $\Lambda $\ to the suite of fundamental physical parameters, which
implies a mass (1) related to Plank's constant of approximately $2\times
10^{-65}$ g and a mass (2) related to the gravitational constant of
approximately $1\times 10^{56}$ g. \ Both can be understood at a deeper
level if the world has more than the 4 dimensions of spacetime. \ In the
prototypical 5D theory, $\Lambda $\ is related to a length which scales the $%
\left( 4+1\right) $ parts of the manifold. \ The latter can most
conveniently be described by the Planck gauge (3) and Einstein gauge (4). \
These for null 5D paths lead to relations (5) and (6), which imply that mass
is quantized.

\section{\protect\underline{Discussion}}

\ \ \ Our main result, that there is a minimum mass of approximately $%
2\times 10^{-65}$ g, raises many questions of both a theoretical and
practical nature. \ For example, if particles of this mass are involved in
interactions, the range of the latter would be large but finite. \ The noted
mass is tiny, even by the standards of particle physics. \ This explains why
mass is apparently unquantized at the levels we have been able to examine,
but also means that a direct test involving current accelerators is
impractical. \ However, the existence of this quantum rests on the
assumption that paths in 5D are null, and this may provide an indirect test
of the approach. \ It is already known that the photons of the cosmic
microwave background have the same temperature to an accuracy of 1 part in $%
10^{5}$, even though according to standard models the parts of the universe
where they originated were out of (4D) causal contact at early times. \ The
conventional way to explain this is, of course, via inflation (an early
period of rapid expansion). \ But it is not clear if this also explains the
uniformity of the properties of massive particles as revealed by the
spectroscopy of remote astronomical sources such as quasars. \ An
alternative view, which needs analysis, is that the universe in 4D appears
uniform because all of its constituents are in causal contact in 5 (or more)
dimensions. \ A related route to testing the approach outlined above
involves work in the laboratory. \ The classical double-slit experiment, and
others like it which show quantum interference, should be revisited, to see
if the apparently baffling behaviour of electrons in ordinary space is due
to the fact that they are in causal contact in higher dimensions.

\ \ \ 

\underline{{\Large Acknowledgements}}

Thanks for comments go to B. Mashhoon and other members of the S.T.M.
consortium. \ This work was supported by N.S.E.R.C.

\ \ \ 

\underline{{\Large References}}

\begin{enumerate}
\item Overduin, J.M., Wesson, P.S. Dark Sky, Dark Matter (Institute of
Physics, London, 2003).

\item Wesson, P.S. Space, Time, Matter (World Scientific, Singapore, 1999).

\item Lineweaver, C.H. Astrophys. J. \underline{505}, L69 (1998).

\item Overduin, J.M. Astrophys. J. \underline{517}, L1 (1999).

\item Wesson, P.S. Phys. Lett. B\underline{276}, 299 (1992).

\item Mashhoon, B., Wesson, P.S., Liu, H. Gen. Rel. Grav. \underline{30},
555 (1998).

\item Randall, L., Sundrum, R. Mod. Phys. Lett. A\underline{13}, 2807 (1998).

\item Arkani-Hamed, N., Dimopoulous, S., Dvali, G.R. Phys. Lett. B\underline{%
429}, 263 (1998).

\item Wesson, P.S., Mashhoon, B., Liu, H., Sajko, W.N. Phys. Lett. B%
\underline{456}, 34 (1999).

\item Youm, D. Phys. Rev. D\underline{62}, 084002 (2000).

\item Seahra, S.S., Wesson, P.S. Gen. Rel. Grav. \underline{33}, 1731 (2001).

\item Youm, D. Mod. Phys. Lett. A\underline{16}, 2731 (2001).

\item Ponce de Leon, J. Mod. Phys. Lett. A\underline{16}, 2291 (2001).

\item Seahra, S.S., Wesson, P.S. Class. Quant. Grav. \underline{20}, 1321
(2003).

\item Kalligas, D., Wesson, P.S., Everitt, C.W.F. Astrophys. J. \underline{%
439}, 548 (1995).

\item Ponce de Leon, J. Gen. Rel. Grav. \underline{20}, 539 (1988).

\item Bellini, M. hep-ph/0206168 (2003).

\item Wesson, P.S. Cosmology and Geophysics \ (Hilger/Oxford Un. Press,
London, 1978).

\item Bridgmann, P.W. Dimensional Analysis (Yale Un. Press, New Haven, 1922).

\item Wesson, P.S. J. Math. Phys. \underline{43}, 2423 (2002).

\item Desloge, E.A. Am. J. Phys. \underline{52}, 312 (1984).

\item Banks, T. hep-ph/0203066 (2002).

\item Wesson, P.S., Liu, H. Int. J. Mod. Phys. D\underline{10}, 905 (2001).

\item Weinberg, S. Rev. Mod. Phys. \underline{52}, 515 (1980).

\item Padmanabhan, T. hep-th/0212290 (2003).

\item Csaki, C., Erlich, J., Grojean, C. hep-th/0012143 (2001).

\item Wesson, P.S. Class. Quant. Grav. \underline{19}, 2825 (2002).
\end{enumerate}

\end{document}